\documentclass[journal]{IEEEtran}

\usepackage{times,amsmath,amssymb,amsfonts,amsthm,graphicx}
\usepackage{cite,color}
\usepackage{bm}
\usepackage{balance}
\usepackage{url}

\def\phibf{\boldsymbol \phi }

\def\abf{{\bf a}}

\def\gbf{{\bf g}}
\def\hbf{{\bf h}}

\def\ubf{{\bf u}}
\def\vbf{{\bf v}}
\def\wbf{{\bf w}}
\def\xbf{{\bf x}}
\def\ybf{{\bf y}}

\def\xbf{{\bf x}}
\def\ybf{{\bf y}}
\def\Abf{{\bf A}}

\def\Cbf{{\bf C}}

\def\Hbf{{\bf H}}

\def\Rbf{{\bf R}}

\def\Gc{{\cal G}}

\def\Kc{{\cal K}}

\def\Pc{{\cal P}}

\def\Sc{{\cal S}}

\def\Xc{{\cal X}}
\def\Yc{{\cal Y}}

\def\ie{{\it i.e.,\ \/}}
\def\nn{\nonumber}
\def\tg{{\tilde{g}}}

\def\tAbf{{\widetilde{\bf A}}}
\def\tCbf{{\widetilde{\bf C}}}
\def\hxbf{{\widehat{\bf x}}}

\def\Re{\mathfrak{R}\mathfrak{e}}
\def\Im{\mathfrak{I}\mathfrak{m}}

\theoremstyle{definition}
\newtheorem{theorem}{Theorem}

\begin{document}

\title{\huge Fast First-Order Algorithm for  Large-Scale Max-Min Fair Multi-Group Multicast Beamforming}

\author{Chong~Zhang,~\IEEEmembership{Student~Member,~IEEE,}
        Min~Dong,~\IEEEmembership{Senior~Member,~IEEE,}
        and~Ben~Liang,~\IEEEmembership{Fellow,~IEEE}%
\thanks{This work was supported by the Natural Sciences and Engineering Research Council of Canada under Discovery
Grants. \textit{(Corresponding author: Min Dong.)}
Chong Zhang and Ben Liang are with the Department of Electrical and Computer Engineering, University of Toronto, Toronto,
ON M5S 1A1, Canada (e-mail: chongzhang@ece.utoronto.ca; liang@ece.utoronto.ca).
Min Dong is with the Department of Electrical, Computer and Software Engineering, Ontario Tech University, Oshawa,
ON L1G 0C5, Canada (e-mail: min.dong@ontariotechu.ca).} }

\maketitle

\begin{abstract}
We propose a first-order fast algorithm for the weighted max-min fair (MMF)  multi-group multicast beamforming
problem in large-scale systems.
Utilizing the optimal multicast beamforming structure  obtained recently,
we convert the nonconvex MMF problem into a  min-max weight minimization problem and show that it is a weakly convex problem. We  propose using the projected subgradient algorithm (PSA)
 to  solve the problem directly, instead of the conventional  method that requires iteratively solving its inverse problem. We show that PSA for our problem has closed-form updates and thus is computationally cheap.
Furthermore,  PSA converges to a near-stationary point of our problem within finite time.
Simulation results show that our PSA-based algorithm offers near-optimal performance with considerably lower computational complexity than existing methods for large-scale systems.
\end{abstract}

\vspace{-0.3em}


\IEEEpeerreviewmaketitle

\vspace{-1.42em}

\section{Introduction}

Content distribution  through wireless multicasting has become increasingly popular  among   wireless applications. Efficient  transmission techniques via  multicast beamforming have become crucial to support high-speed content distribution. With  massive multiple-input multiple-output (MIMO) becoming the essential technology  for future networks, it is critical to develop effective and computationally efficient multicast beamforming solutions suitable for  large-scale systems.

Early works studied the multicast beamforming  design for  traditional multi-antenna systems in various scenarios, including a single user group or multiple user groups~\cite{Sidiropoulos&etal:TSP2006,Karipidis&etal:TSP2008}, multi-cell networks~\cite{Xiang&Tao&Wang:IJWC:13}, and relay networks~\cite{DongLiang:CAMSAP13}.
Since the family of multicast beamforming problems are nonconvex and NP-hard, the existing works have focused on developing numerical algorithms or signal processing techniques for good suboptimal solutions. Semi-definite relaxation (SDR)  has been a widely adopted common approach~\cite{Sidiropoulos&etal:TSP2006,Karipidis&etal:TSP2008,Xiang&Tao&Wang:IJWC:13}. However, as wireless systems are becoming large-scale, the successive convex approximation (SCA) method \cite{Marks&Wright:OperReas1978}  becomes more  popular for its computational and performance advantages over  SDR as the size of the problem grows~\cite{Tran&etal:SPL2014,Mehanna&etal:2015,Christopoulos&etal:SPAWC15}.
Despite the  improvement,   SCA  relies on second-order interior-point methods (IPMs) to solve each convex approximation,  where the computational complexity   is still  too high   for massive MIMO systems.
 Several  algorithms were proposed to improve the computational efficiency at each SCA iteration,
 such  as zero-forcing pre-processing~\cite{Sadeghi&etal:TWC17} and alternating direction method of multipliers (ADMM)~\cite{Chen&Tao:ITC2017} for  multi-group scenarios, and  first-order methods~\cite{Konar&Sidiropoulos:TSP2017} for single-group scenarios.
The optimal multicast beamforming structure has been  obtained recently in \cite{Dong&Wang:TSP2020}, which is shown to be a weighted minimum mean square
error (MMSE) filter with an inherent low-dimensional structure.  This  structure helps convert  the beamforming  problem into a weight optimization problem of a much lower dimension \cite{Dong&Wang:TSP2020}, allowing design opportunities for efficient algorithms  for massive MIMO systems.

The multi-group multicast beamforming design can be cast into either a quality-of-service (QoS) problem for power minimization with signal-to-interference-and-noise (SINR) guarantees, or a max-min fair (MMF) problem for maximizing the minimum SINR subject to some transmit power budget. They are inverse problems~\cite{Karipidis&etal:TSP2008}. Although both problems are nonconvex, the MMF problem is a max-min problem that is  much more complicated to solve than the QoS problem~\cite{Sidiropoulos&etal:TSP2006,Karipidis&etal:TSP2008}.   Typically, the solution to the  MMF problem is obtained via iteratively solving its inverse QoS problem along with a bi-section search~\cite{Karipidis&etal:TSP2008, Christopoulos&etal:SPAWC15, Chen&Tao:ITC2017, Dong&Wang:TSP2020}.
The QoS problem at each iteration can then be solved by either  SDR or SCA.
This additional layer of  iteration leads to high computational complexity, especially for large-scale systems.

To address the above issue, in this letter, we propose a fast first-order algorithm for the weighted MMF    multi-group multicasting problem.
We focus on the min-max weight optimization problem, which is transformed from the original MMF problem by using the optimal beamforming structure  \cite{Dong&Wang:TSP2020}. We show that this converted problem is weakly convex, and the  projected subgradient algorithm (PSA) \cite{Polyak:book1987} can be efficiently used to solve it  \emph{directly}.
In particular, we show that  for our problem, PSA provides closed-form  subgradient update and projection and thus, is computationally cheap.
Furthermore, based on the recent  convergence result  for weakly convex problems, we show  that  PSA    converges to a near-stationary point of our problem within finite time.
 We further propose an initialization method  for  faster convergence. Our simulation results show that our PSA-based algorithm offers  near-optimal performance with substantially lower computational complexity than the existing state-of-the-art algorithms for large-scale systems.

\allowdisplaybreaks

\vspace{-.9em}
\section{Problem Formulation}
We consider a downlink multi-group multicast beamforming scenario, where the base station (BS) equipped with $N$ antennas transmits messages to $G$ multicast groups. Each group receives a common message  that is independent of the messages to other groups. Denote the set of group indices by $\Gc \triangleq \{1,\cdots,G\}$.
Assume that there are $K_i$ single-antenna users in group $i$, with the set of user indices denoted by $\mathcal{K}_{i} \triangleq \{1,\cdots,K_i\}, \, i\in\Gc$. The total number of users in all groups is denoted by $K_{\text{tot}}\triangleq \sum_{i=1}^{G}K_i$.
Let $\wbf_i\in \mathbb{C}^{N}$ be the multicast beamforming vector for group $i$, and let $\hbf_{ik}\in \mathbb{C}^{N}$ be the channel vector from the BS to  user $k$ in group $i$, for $k\in \Kc_i$, $i\in \Gc$. The received signal at user $k$ in group $i$ is given by \vspace*{-.7em}
\begin{align*}
y_{ik}={\bf{w}}^{H}_{i}{\bf{h}}_{ik}s_{i}+\sum_{j\neq i}{\bf{w}}^{H}_{j}{\bf{h}}_{ik}s_{j}+n_{ik}
\end{align*}
where $s_i$ is the  symbol transmitted to group $i$ with $E[|s_{i}|^{2}] = 1$, and $n_{ik}$ is the receiver additive white Gaussian noise with zero mean and variance $\sigma^{2}$. The received SINR at this user is \begin{align}
\text{SINR}_{ik}=\frac{|\wbf_i^{H}\hbf_{ik}|^{2}}{\sum_{j\neq i}|{\bf{w}}_{j}^{H}\hbf_{ik}|^{2}+\sigma^2}.
\end{align}
The BS transmit power is given by $\sum_{i=1}^{G} \|\wbf_i\|^2$.

This letter focuses on the weighted MMF multicast beamforming problem, \ie  maximizing the minimum weighted SINR among all users, subject to the BS transmit power constraint. We assume that all $\hbf_{ik}$'s are perfectly known at the BS.
Define ${\bf{w}}\triangleq[{\bf{w}}_{1}^H, \cdots, {\bf{w}}_{G}^H]^H$. The weighted MMF problem is given by\vspace*{-.5em}
\begin{align}
\Sc_o: \; \max_{{\bf{w}}} \min_{i,k} &\,\, \frac{\text{SINR}_{ik}}{\gamma_{ik}}
\quad \text{s.t.}  \;\; \sum_{i=1}^{G} \|\wbf_i\|^2\leq P \nn
\end{align}
where $P$ is the maximum power budget at the BS, and  $\{\gamma_{ik}\}$ are the weights to control the grade of service or fairness among users.

Problem $\Sc_o$ is a nonconvex max-min optimization problem and is known to be NP-hard. Existing methods in the literature are through iteratively solving the dual problem of   $\Sc_o$  -- the QoS problem, \ie minimizing the transmit power subject to minimum SINR targets \cite{Karipidis&etal:TSP2008, Christopoulos&etal:SPAWC15}.
Specifically, consider the following equivalent problem to  $\Sc_{o}$: \vspace*{-.3em}
\begin{align}
\Sc_1: & \;\; \max_{{\bf{w}},t} \; t \nn \\[-1em] \text{s.t.} & \;\; \text{SINR}_{ik}\geq t\gamma_{ik}, k\in \Kc_i, i\in \Gc, \quad  \sum_{i=1}^{G} \|\wbf_i\|^2\leq P. \nn
\end{align}
 The dual QoS problem to $\Sc_1$ is given as follows: \vspace{-.3em}
\begin{align}
\Pc_o: & \min_{{\bf{w}}} \,\, \sum_{i=1}^{G} \|\wbf_i\|^2  \quad \text{s.t.} \quad \text{SINR}_{ik}\geq t\gamma_{ik},\,\, k\in \mathcal{K}_i, i\in \Gc. \nn
\end{align}
  The solution to $\Sc_{o}$ is computed by solving $\Pc_{o}$  along with a bi-section search over $t$ until the transmit power is equal to $P$.
The popular methods in the literature to solve the nonconvex problem   $\Pc_{o}$ are  SDR and, recently, SCA. SCA has an advantage in both performance and computational efficiency for  large-scale problems. It convexifies the problem first and relies on the second-order IPM to solve the corresponding convex approximation problem \cite{Karipidis&etal:TSP2008, Christopoulos&etal:SPAWC15, Dong&Wang:TSP2020}. However, the    computational complexity of the IPM is still high for large-scale problems. As a result, the iterative method to solve $\Sc_o$ via $\Pc_o$  incurs high computational complexity for wireless systems with large-scale antenna arrays or a large number of users.

In this letter, we propose a fast first-order algorithm to  solve $\Sc_{o}$ directly with low computational complexity.

\vspace{-.5em}
\section{Preliminary: Optimal Multicast Beamforming Structure}

The  structure of the optimal multicast beamforming solution  to $\Sc_{o}$ has recently been obtained in \cite{Dong&Wang:TSP2020}, which is shown to be a weighted MMSE filter. Under this solution structure,  $\Sc_{o}$ is transformed into an equivalent weight optimization problem
of a much smaller size that is independent of the number of antennas $N$. This structure can be utilized for substantial computational saving for a solution in large-scale systems.
The optimal multicast beamforming solution is given by \cite[Theorem 2]{Dong&Wang:TSP2020}
\vspace*{-.5em}
\begin{align}
\wbf^{o}_{i} = \Rbf^{-1}\Hbf_{i}\abf^{o}_{i},\quad i\in \Gc
\label{asymptotic_structure}
\end{align}
where $\Hbf_{i} \triangleq [\hbf_{i1}, \cdots, \hbf_{iK_{i}}]$ is the channel matrix for group $i$,
 $\abf^{o}_{i}\in \mathbb{C}^{K_i}$ is the optimal weight vector for group $i$, and $\Rbf$ is the (normalized) noise plus weighted channel covariance  (of all users)  matrix given in a semi-closed form. To further simplify the required computation, the  approximate expression of  $\Rbf$  is  obtained in \cite{Dong&Wang:TSP2020}. Express each channel  as $\hbf_{ik}=\sqrt{\beta_{ik}}\gbf_{ik}$, where $\beta_{ik}$ is the
channel variance,  and $\gbf_{ik}$ is the normalized channel vector representing the small-scale fading whose elements are i.i.d. zero mean.
The  approximate expression of $\Rbf$, for large $N$, is given by\vspace{-.3em}
\begin{align}
\widetilde{\Rbf} = {\bf{I}}+\frac{P}{\sigma^{2}}\sum_{i=1}^{G}\sum_{k=1}^{K_{i}}\frac{\eta_{ik}}{ \sum_{i'=1}^{G}\!\sum_{k'=1}^{K_{i'}}\!\frac{\eta_{i'\!k'}}{\beta_{i'\!k'}}}\gbf_{ik}\gbf_{ik}^{H}
\label{eq_expression_R_general}
\end{align}
where $\eta_{ik} \triangleq \frac{\gamma_{ik}}{{(N-\mathop{\sum\sum}\limits_{jl\neq ik}\gamma_{jl})}}$.
In particular, for $\gamma_{ik} = \gamma$, $\forall i, k$, $\widetilde{\Rbf}$ in (\ref{eq_expression_R_general}) is further simplified to\vspace{-.5em}
\begin{align}
\widetilde{\Rbf} = {\bf{I}}+\frac{P\bar{\beta}}{\sigma^{2}K_{\text{tot}}}\sum_{i=1}^{G}\sum_{k=1}^{K_{i}}\gbf_{ik}\gbf_{ik}^{H}
\label{eq_asymp_R_common_gamma}
\end{align}
where $\bar{\beta} \triangleq \frac{K_{\text{tot}}}{\sum_{i=1}^{G}\!\sum_{k=1}^{K_{i}}\!\!\frac{1}{\beta_{ik}}}$
is the harmonic mean of the channel variances of all users.
With the  solution $\wbf^o_{i}$ in (\ref{asymptotic_structure}) and  $\widetilde\Rbf$ to approximate $\Rbf$,
the original problem $\mathcal{S}_{o}$ can be transformed into the following
 weight optimization problem\vspace{-.2em}
\begin{align*}
\Sc_{2}:   \;\max_{\abf} \min_{i,k} \;\;& \frac{1}{\gamma_{ik}}\frac{\abf_{i}^{H}\tAbf_{iik}\abf_{i}}{\sum_{j\neq i}\abf_{j}^{H}\tAbf_{jik}\abf_{j}+\sigma^2}  \nonumber\\
   \text{s.t.}\;\; & \sum_{i=1}^{G} \|\tCbf_{i}\abf_{i}\|^2\leq P
\end{align*}
where $\abf\triangleq[\abf_{1}^H, \cdots, \abf_{G}^H]^H$,
$\tCbf_{i} \triangleq \widetilde{\Rbf}^{-1}\Hbf_{i}$, and $\tAbf_{jik} \triangleq \tCbf^{H}_{j}\hbf_{ik}\hbf^{H}_{ik}\tCbf_{j}$, $k\in\Kc_{i}$, $i,j\in\Gc$.
Note that the dimension of weight vector $\abf$ ($K_\text{tot}$) is much lower
 than that of the beamforming vector $\wbf$ ($GN$) in massive MIMO systems with $K_i \ll N$.
 Hence,  the  complexity in computing the beamforming solution is substantially reduced by optimizing  $\abf$  in $\Sc_2$, instead of $\wbf$ in $\Sc_o$.

Note that  \cite{Dong&Wang:TSP2020} focuses on obtaining the optimal beamforming structure, while it still adopts the commonly used numerical algorithms, such as SDR or SCA, when solving the optimization problems.   Different from \cite{Dong&Wang:TSP2020}, in this letter,
we utilize the optimal structure and  focus on proposing a fast numerical algorithm for solving the max-min  optimization problem, which provides a more efficient computational method to obtain a solution to the MMF problem.

\vspace{-.5em}
\section{First-Order Fast Algorithm}
Using the optimal beamforming structure,
in this section, we propose a fast first-order algorithm to solve $\Sc_2$.  Problem   $\Sc_{2}$ is a nonconvex max-min problem. Based on the  structure of $\Sc_2$, we show that    PSA can be applied to compute a near-stationary solution to $\Sc_2$  efficiently.

\vspace{-1em}
\subsection{Problem Reformulation}

For the purpose of computation, we express all complex quantities in $\Sc_2$ using their real and imaginary parts.
Define $\xbf_{i}\triangleq [\Re{\{\abf_{i}\}}^T, \Im{\{\abf_{i}\}}^T]^T$,
$\Cbf_{i}\triangleq
\begin{bmatrix}
\Re{\{\tCbf_{i}\}} & \!\!\! -\Im{\{\tCbf_{i}\}} \\
\Im{\{\tCbf_{i}\}} &  \!\!\!\Re{\{\tCbf_{i}\}}
\end{bmatrix}$,
$\Abf_{jik}\triangleq
\begin{bmatrix}
\Re{\{\tAbf_{jik}\}} & \!\!\! -\Im{\{\tAbf_{jik}\}} \\
\Im{\{\tAbf_{jik}\}} & \!\!\! \Re{\{\tAbf_{jik}\}} \nn
\end{bmatrix}$,
for $k\in\Kc_{i}$, $i,j\in\Gc$. It follows that $\|\tCbf_{i}\abf_{i}\|^{2} = \|\Cbf_{i}\xbf_{i}\|^2$
and
$\abf_{j}^{H}\tAbf_{jik}\abf_{j} = \xbf_{j}^{T}\Abf_{jik}\xbf_{j}$.
Using the above, we can express problem $\mathcal{S}_2$ equivalently  in the real domain
as \vspace{-.4em}
\begin{align*}
\Sc_{3}: \;  \max_{\xbf\in\Xc}\min_{i,k} \frac{1}{\gamma_{ik}}\frac{\xbf_{i}^{T}\Abf_{iik}\xbf_{i}}{\sum_{j\neq i}\xbf_{j}^{T}\Abf_{jik}\xbf_{j}+\sigma^{2}}
\end{align*}
where $\xbf\triangleq[\xbf_1^T,\ldots,\xbf_G^T]^T$, and $\Xc\triangleq \{\xbf: \sum_{i=1}^{G}\|\Cbf_{i}\xbf_{i}\|^2\leq P\}$ is the compact convex feasible set of $\mathcal{S}_{3}$.
Define $\phi_{ik}(\xbf) \triangleq \frac{-\xbf_{i}^{T}\Abf_{iik}\xbf_{i}}{\gamma_{ik}(\sum_{j\neq i}\xbf_{j}^{T}\Abf_{jik}\xbf_{j}+\sigma^{2})}$,  $k\in \Kc_{i}, i\in \Gc$. Then, we can rewrite  $\Sc_3$ in an equivalent min-max form as  $\min_{\xbf\in\Xc}\max_{i,k} \phi_{ik}(\xbf)$,
which is further  equivalent to\vspace*{-.5em}
\begin{align}
\Sc_4:\;  \min_{\xbf\in\Xc}\max_{\ybf\in\Yc}\,\, f(\xbf, \ybf)\nn
\end{align}
where $f(\xbf, \ybf)\triangleq \phibf^{T}(\xbf)\ybf$, with $\phibf(\xbf)\in\mathbb{R}^{K_{\text{tot}}}$ containing all $\phi_{ik}(\xbf)$'s, and  $\Yc \triangleq \{\ybf: \ybf\succcurlyeq {\bf 0}, {\bf 1}^{T}\ybf = 1\}$ is a probability simplex, which is  a compact  convex set.\footnote{Note that by introducing $\ybf$, we transform $\Sc_{3}$ into $\Sc_{4}$. The structure in $\Sc_{4}$ will benefit our exposition in Section~\ref{sec_PSA} for the proposed algorithm and convergence analysis.}
An optimal solution to the inner maximization of $\Sc_4$ is $\ybf=[0,\ldots,0,1,0,\ldots,0]^T$, with $1$  at some $i$th position.
Note that $f(\xbf, \ybf)$  is concave in $\ybf$ and nonconvex in\ $\xbf$.
Thus, $\Sc_4$ is a nonconvex-concave min-max problem and is NP-hard.
Let $g(\xbf) \triangleq \max_{\ybf\in\Yc} f(\xbf, \ybf)$.
Then, we express $\Sc_4$ as \vspace*{-.5em}
\begin{align*}
\Sc_5: \;\min_{\xbf\in\Xc}\,\, g(\xbf).\nn
\end{align*}
Note that $g(\xbf)$ is nonconvex. If $g(\xbf)$ is differentiable,
one can use the projected gradient descent \cite{Levitin&Polyak:USSR1966} to solve $\Sc_{5}$.
However, in our problem, $g(\xbf)$ may not be differentiable,
and its gradient $\nabla g(\xbf)$ may not exist.
In what follows, by examining the structure of the problem, we propose to use  PSA \cite{Polyak:book1987} to find
a solution at the vicinity of a stationary point for $\Sc_{5}$.

\subsection{The Projected Subgradient Algorithm}
\label{sec_PSA}
 We first show the  structure of our problem.
We assume the channel gain is finite for each user:    $\|\hbf_{ik}\| < \infty$, $\forall k,i$. Thus, all elements in $\Abf_{jik}$ are  finite,  $\forall k\in\Kc_{i}$, $i,j\in\Gc$. Also, note that $\gamma_{ik} > 0$, $\sigma^2 > 0$. It follows that,  since $\Xc$ and $\Yc$ are compact,
the gradient $\nabla_{\xbf} f(\xbf, \ybf)$ is finite for any $\xbf\in\Xc, \ybf\in \Yc$. Thus, there exists a constant $L > 0$, such that
$\|\nabla_{\xbf} f(\xbf_{1}, \ybf) - \nabla_{\xbf} f(\xbf_{2}, \ybf)\| \leq L \|\xbf_{1} - \xbf_{2}\|$, for any $\xbf_{1},\xbf_{2} \in \Xc$, $\ybf\in\Yc$. This means that, $f(\xbf, \ybf)$ is an $L$-smooth function of $\xbf\in\Xc$,
which satisfies the following  \cite{Beck:book2017} \vspace{-.3em}
\begin{align}
f(\xbf_{2}, \ybf) \geq & f(\xbf_{1}, \ybf) + \nabla_{\xbf}f(\xbf_{1}, \ybf)^T(\xbf_{2} - \xbf_{1}) \nn\\
& - \frac{L}{2}\|\xbf_{2} - \xbf_{1}\|^{2}, \,\, \forall \, \xbf_{1},\xbf_{2} \in \Xc, \, \forall \, \ybf \in \Yc.
\label{weaker_conv}
\end{align}
Next, we show that  $\nabla_{\xbf}f(\xbf, \ybf)$   is a subgradient of $g(\xbf)$. The Fr$\acute{\text{e}}$chet subdifferential of $g(\xbf)$ is the set of  subgradients of $g(\xbf)$ defined by
$\partial g(\xbf) \triangleq \{\vbf \, | $ ${\liminf}_{\xbf^{\prime}\rightarrow \xbf}$  $\frac{g(\xbf^{\prime}) - g(\xbf) - \vbf^{T}(\xbf^{\prime} - \xbf)}{\|\xbf^{\prime} - \xbf\|} \geq 0\}$ \cite{Rockafellar&Wets:book2009}.
By the definition of $g(\xbf)$ and from \eqref{weaker_conv},
for any  $\xbf^{\prime}\in\Xc$,
we have \vspace{-.6em}\begin{align}
g(\xbf^{\prime})
&  \geq f(\xbf, \ybf)  + \nabla_{\xbf}f(\xbf, \ybf)^{T}(\xbf^{\prime} - \xbf) - \frac{L}{2}\|\xbf^{\prime} - \xbf\|^{2}  \nn\\
& = g(\xbf)\!+\! \nabla_{\xbf}f(\xbf, \ybf)^T(\xbf^{\prime} - \xbf) \!-\! \frac{L}{2}\|\xbf^{\prime} - \xbf\|^{2}.
\label{eq_g_ineq}
\end{align}
After rearranging the terms at both sides of the inequality in \eqref{eq_g_ineq}
and taking $\lim\inf$ for $\xbf'\to \xbf$, we conclude that $\nabla_{\xbf}f(\xbf, \ybf)\in \partial g(\xbf)$.

Following the above result, we propose to solve $\Sc_{5}$ by  PSA with the following updating procedure: \vspace{0.5em}

\fbox{%
\begin{minipage}{3.14 in}
At iteration $j$:
\begin{align}
&  \ybf^{(j)}  \in \underset{\ybf\in\Yc}{\arg\max}\,\, f(\xbf^{(j)}, \ybf), \label{update_y}\\
&  \xbf^{(j+1)} = \Pi_{\Xc}\big(\xbf^{(j)} - \alpha \nabla_{\xbf}f(\xbf^{(j)}, \ybf^{(j)})\big) \label{update_x}
\end{align}
\end{minipage}}\vspace{0.5em}
where $\alpha > 0$ is the step size, and
$\Pi_\Xc(\xbf)$ denotes the projection of point $\xbf$ onto set $\Xc$,
given by
\begin{align}
 \Pi_{\Xc}(\xbf)=
\begin{cases}
\sqrt{\frac{P}{P_{\xbf}}}\xbf   & {\xbf\notin\Xc} \\
\xbf           & {\xbf\in\Xc}
\end{cases}
\label{eq_projection}
\end{align}
where $P_{\xbf} \triangleq \sum_{i=1}^{G}\|\Cbf_{i}\xbf_i\|^2$.

Note that the inherent structure of our problem makes  PSA
particularly suitable for solving $\Sc_5$.
First, $\ybf^{(j)}$ in  \eqref{update_y} can be directly obtained by taking the maximum among $\phi_{ik}(\xbf^{(j)})$'s.  Second, the projection $\Pi_{\Xc}(\xbf)$ is a  simple closed-form function in \eqref{eq_projection}, and  $\nabla_{\xbf}f(\xbf, \ybf)$ has a closed-form expression. Thus, the computation of $\xbf^{(j+1)}$ in \eqref{update_x} is inexpensive.  Below, we discuss the convergence result for the proposed PSA.

\subsubsection{Convergence Analysis}Based on the recent results on weakly convex problems  \cite{Jin&etal:PMLR2020,Damek&Dmitriy:SIAM2019,Chen&etal:TAC2021}, we show  that PSA converges within  finite time to a near-stationary point of $\Sc_5$.
Recall that $f(\xbf, \ybf)$ is $L$-smooth over $\Xc$,
and  $\Yc$ is compact. It follows that
  $g(\xbf) = \max_{\ybf\in\Yc} f(\xbf, \ybf)$ is
\emph{$L$-weakly convex} over $\Xc$, \ie $g(\xbf) + \frac{L}{2}\|\xbf\|^{2}$ is convex for $\xbf\in\Xc$ \cite[Lemma 1]{Kiran&etal:NIPS2019}.
Consider the extension of  $g(\xbf)$ to $\mathbb{R}^{2K_\text{tot}}$: $\tilde{g}(\xbf)= g(\xbf) + \mathbb{I}_{\Xc}(\xbf) $, where $\mathbb{I}_{\Xc}(\xbf)$ is an indicator function, taking   $0$ if $\xbf\in\Xc$ and $\infty$ otherwise.  Define the Moreau envelope \cite{Rockafellar:book2015} of $\tilde{g}(\xbf)$   as
\begin{align} \label{Moreau}
\tilde{g}_{\lambda}(\xbf) \triangleq \min_{\xbf'\in \mathbb{R}^{2K_\text{tot}}} \{\tilde{g}(\xbf') + \frac{1}{2\lambda} \|\xbf' - \xbf\|^{2}\}
\end{align}
where $\lambda < 1/L$. The Moreau envelope $\tilde{g}_{\lambda}(\xbf)$ is a smooth approximation
to the non-smooth but $L$-weakly convex function $g(\xbf)$ over $\Xc$. Note from the earlier discussion that  the objective function of the minimization problem in  \eqref{Moreau} is strictly convex. Let $\hxbf\triangleq \arg\min_{\xbf'}\{\tilde{g}(\xbf') + \frac{1}{2\lambda} \|\xbf' - \xbf\|^{2}\}$.  Then, we have $\hxbf \in \Xc$ and \vspace{-.5em}
\begin{align}\label{Moreau_x}
\|\hxbf - \xbf\|=\lambda\|\nabla{\tilde{g}_{\lambda}(\xbf)}\|.
\end{align}
Thus, $\|\nabla{\tilde{g}_{\lambda}(\xbf)}\|\leq \epsilon$ implies that \cite{Rockafellar:book2015}
\vspace{-.5em}\begin{align}
\|\hxbf - \xbf\| \le \lambda\epsilon, \quad \text{and} \quad
\underset{\ubf\in\partial \tilde{g}(\hxbf)}{\min} \|\ubf\| \leq \epsilon.
\label{weak_gradient}
\end{align}
The above  means that a small gradient $\|\nabla{\tilde{g}_{\lambda}(\xbf)}\|\leq \epsilon$ implies that $\xbf$ is close to  a point  $\hxbf$ that is a near-stationary (\ie $\epsilon$-stationary) point   of $\Sc_{5}$.
Hence,  $\|\nabla{\tilde{g}_{\lambda}(\xbf)}\|$ provides a near-stationarity
measure of $\xbf$ to a stationary point  of $\Sc_5$.
Based on this, we have the  convergence result  of  PSA for $\Sc_5$ below.\vspace{-.3em}
 \begin{theorem}\label{thm1}
 Assume the continuous function $f(\xbf,\ybf)$ is $C$-Lipschitz
 over $\Xc \times \Yc$.
Define
$D$ $\triangleq$ $\max_{\xbf_{1},\xbf_{2}\in\Xc}$ $\|\xbf_{1}-\xbf_{2}\|$,
$M \triangleq$ $\max_{\xbf\in\Xc,\ybf\in\Yc}$ $\|\nabla_{\xbf}f(\xbf, \ybf)\|$,
and
$\Delta\triangleq \min{\{LD^{2}, CD\}}$.
Starting from $\xbf^{(0)} \in \Xc$, let $J$ be the total number of iterations in PSA.
 Let step size $\alpha = \sqrt{\frac{\Delta}{LM^{2}(J+1)}}$.  Let the output of PSA be $\bar{\xbf}=\xbf^{(j)}$, where $j\sim\text{Uniform}[0,J]$. Then,  $\bar{\xbf}$  satisfies
 \begin{align}
E\|\nabla{\tg_{\frac{1}{2L}}(\bar{\xbf})}\|^{2}\leq \frac{4\sqrt{\Delta L M^{2}}}{\sqrt{J+1}}.
  \label{ineq_avrg_epsilon}
 \end{align}
\end{theorem}
\IEEEproof
See Appendix~\ref{appA}.
\endIEEEproof
Theorem~\ref{thm1}  indicates that if we take a random sample in $\{\xbf^{(j)}\}_{j=0}^J$ as the output of PSA $\bar{\xbf}$, then $E\|\nabla{\tg_{\frac{1}{2L}}(\bar{\xbf})}\|^{2}$ decreases in the order of (at most)  $O(\frac{1}{\sqrt{\!J}})$. A more direct way to interpret this result is that, to obtain the output  $\bar{\xbf}$ satisfying $E\|\nabla{\tg_{\frac{1}{2L}}(\bar{\xbf})}\|^{} \le \epsilon$, the required number of iterations $J$ for PSA is \emph{at most} $O(\epsilon^{-4})$.
Thus, for weakly convex problem $\Sc_{5}$,  Theorem~\ref{thm1} shows that PSA converges  within  finite time, upper bounded by $O(\epsilon^{-4}),$ to an  $\epsilon$-accuracy point of $\Sc_5$.

By Theorem~\ref{thm1}, we can also set the stopping criterion for PSA. The convergence analysis in  Theorem~\ref{thm1} is based on a random sample in $\{\xbf^{(j)}\}_{j=0}^J$. Thus, to implement PSA, we  can  set a random stopping point $J^o < J$ for PSA and use   $\xbf^{(J^o)}$ as the algorithm output.
\subsubsection{Initialization}
An  easy-to-compute good initial point is essential to accelerate the convergence of PSA.  Using \eqref{asymptotic_structure} to transfer  $\Pc_{o}$ into optimizing $\abf$ (denoted by $\Pc_o'$) instead of $\wbf$,
we propose to use SDR with  Gaussian randomization (GR)  \cite{Dong&Wang:TSP2020} to solve $\Pc_o'$
along with one bi-section search  over $t$ to generate the initial point.
 The one-step bi-section  is inexpensive and is intended to find  $\xbf^{(0)}$ that is closer to the optimal solution.
Note that this initial point  may not be feasible in $\Xc$. Nonetheless,
after one iteration  via the projection step in \eqref{update_x}, the subsequent points are feasible. This initialization method has low computational complexity and generates an initial point very close to a stationary point when $\Sc_5$ is of a small to moderate size.

\section{Simulation Results}
\label{sec:simulations}
 We set $G=3$,  $K_{i} = K$,
and  SINR target   $\gamma_{ik} = \gamma = 10~\text{dB}$, $\forall k, i$.
The channels are generated i.i.d.\ as $\hbf_{ik}\sim\mathcal{CN}({\bf{0}},{\bf{I}})$, and
the receiver noise variance is $\sigma^{2} = 1$.
We set  $P/\sigma^2=10~\text{dB}$. The approximate expression $\widetilde \Rbf$ in (\ref{eq_asymp_R_common_gamma}) is used in the simulation. For PSA,
we set the step size $\alpha = 0.01$ and set the stopping criterion as $|g(\xbf^{(j+1)})-g(\xbf^{(j)})|\le 10^{-5}$. Based on various simulation experiments, PSA generally converges within 30 $\sim$ 5000 iterations.\footnote{We have studied  different values of $\alpha$ and found  $\alpha = 0.01$ generally provides suitable trade-off between performance and convergence speed.} Besides our proposed PSA with the initialization method, we consider the following methods for comparison:
1) The upper bound for $\Sc_{o}$: It is obtained by solving $\Pc_o$ using SDR along with the bi-section search over $t$;
2) SDR: It uses the optimal structure in \eqref{asymptotic_structure}
with  $\widetilde \Rbf$ in (\ref{eq_asymp_R_common_gamma}) and solves $\Sc_o$ by solving  $\Pc_o$ using SDR plus GR along with the bi-section search over $t$   \cite{Dong&Wang:TSP2020};
3) SCA: It uses the optimal structure in \eqref{asymptotic_structure}
with   $\widetilde \Rbf$ in (\ref{eq_asymp_R_common_gamma}) and solves $\Sc_o$ by solving  $\Pc_o$ via SCA and the bi-section search over $t$.  CVX is used  in each SCA iteration   \cite{Dong&Wang:TSP2020}.
The proposed initialization method is used in all methods.

\begin{figure}[t]
\centering
\includegraphics[scale=.44]{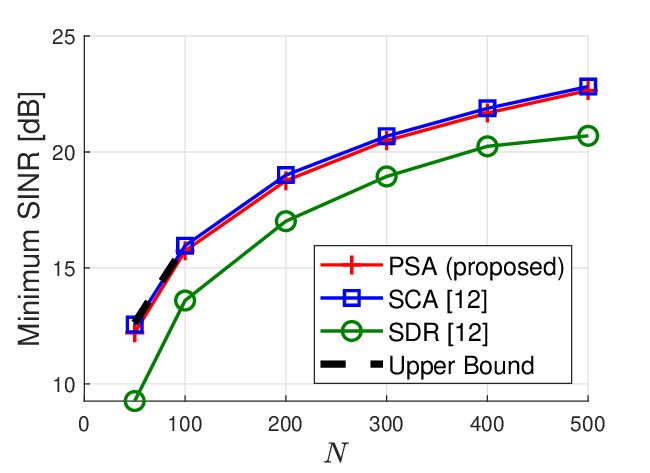}\hspace*{-1em}\includegraphics[scale=0.44]{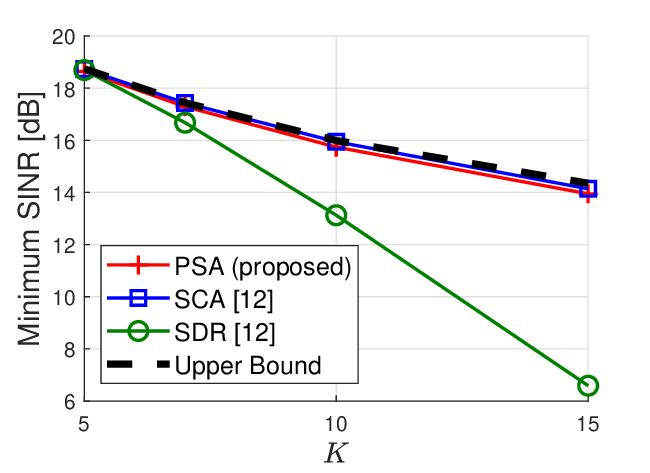}
\caption{Left: Average minimum SINR vs. $N$ ($G = 3$,\,$K = 10$). Right: Average minimum SINR vs. $K$ ($G = 3$,\,$N = 100$).}
\label{Fig:Mini_SINR_N} \vspace*{-1em}
\end{figure}

Fig.~\ref{Fig:Mini_SINR_N}-Left shows the average minimum SINR
vs.\  $N$  for $K = 10$.
Both PSA and SCA nearly attain the upper bound for all values of $N$.  SDR is about $2$dB worse, as the approximation by  SDR
deteriorates when the number of constraints ($GK$) becomes large. Compared with other methods, our proposed PSA is much more computationally efficient in obtaining the solution.
 Table \ref{Table:Compt_time_N} shows the corresponding computation time, which includes computing the initial point. The average computation time of PSA
is only about $3\%$ of that of SCA and about $20\%$ of that of SDR.

Fig.~\ref{Fig:Mini_SINR_N}-Right shows the average minimum SINR vs.\ $K$  for  $N = 100$.
Again, both  PSA and SCA nearly attain the upper bound for all values of $K$, while SDR deteriorates substantially as $K$ becomes large.
The corresponding average computation times (including the initial point)   are shown in Table \ref{Table:Compt_time_K}, which again show
that our proposed PSA is a fast algorithm with much lower computational complexity than SCA and SDR.

\begin{table}[t]
\footnotesize
 \renewcommand{\arraystretch}{1.1}
\centering
\caption{\footnotesize Average Computation Time over $N$ (sec.) ($G = 3$, $K = 10$)}
\vspace{-1em}
\begin{tabular}{l|c|c|c|c|ccc}
\hline
\hspace{3em}$N$            &  100  &  200  &  300  & 400   &  500                             \\ \hline\hline
PSA (proposed)         &  1.830  &  2.031  &   2.524  &   2.753  &   2.821                \\ \hline
SCA \cite{Dong&Wang:TSP2020}   &  87.12  &  81.17  &  81.71  &  86.63  & 90.54           \\ \hline
SDR \cite{Dong&Wang:TSP2020}       &  11.47   &  12.36  &  13.07  &  14.41  & 14.90           \\\hline
\end{tabular}
\label{Table:Compt_time_N}
\vspace{1em}
\footnotesize
\centering
\caption{\footnotesize Average Computation Time over $K$ (sec.) ($G = 3$, $N = 100$)}
\vspace{-1em}
\begin{tabular}{l|c|c|c|c}
\hline
\hspace{3em} $K$            &   5    &    7   &   10   & 15                            \\ \hline\hline
PSA (proposed)   &  0.814  &    1.128  &    1.851  &   3.676                       \\ \hline
SCA \cite{Dong&Wang:TSP2020} &   24.54  &   47.20  &   87.53   &  174.1           \\ \hline
SDR  \cite{Dong&Wang:TSP2020}     &   6.063  &   8.342  &   11.33   &  21.60           \\ \hline
\end{tabular}\vspace{-1em}
\label{Table:Compt_time_K}
\end{table}

\section{Conclusion}
In this letter, we have proposed a fast algorithm for  multi-group multicast  MMF beamforming using the optimal beamforming structure. We have shown that the nonconvex MMF problem can be transformed into an $L$-weakly convex optimization problem,  which we have proposed using   PSA to  solve directly.  Under our problem structure, PSA yields a closed-form updating procedure that is highly computationally inexpensive. We provide the convergence result to the proposed PSA.
Simulation results demonstrate that PSA  provides a near-optimal performance with a substantially lower computational complexity than the existing algorithms for large-scale systems.

\vspace{-.1em}

\appendices
\section{Proof of Theorem~\ref{thm1}}\label{appA}
\IEEEproof
Our proof   follows the  proof techniques  of Theorem 3.1 in \cite{Damek&Dmitriy:SIAM2019}.\footnote{The convergence analysis in \cite{Damek&Dmitriy:SIAM2019} is for a proximal stochastic subgradient method   for stochastic optimization of weakly convex   functions.  Since PSA is different from the stochastic method, that convergence result cannot be directly used.} Let $\hxbf^{(j)} \triangleq \arg\min_{\xbf}\{\tilde{g}(\xbf) + L \|\xbf - \xbf^{(j)}\|^{2}\}$.
 Based on $\tilde{g}_\lambda(\xbf)$ in  \eqref{Moreau}, with $\lambda = \frac{1}{2L}$,  we have\vspace{-.3em}
\begin{align}
& \hspace*{-.5em}\tg_{\!\frac{1}{2L}}(\xbf^{(j+1)})   \le  \tg(\hxbf^{(j)})\! + \! L\|\hxbf^{(j)} \!- \! \xbf^{(j+1)}\|^2 \nn \\
& \hspace*{-.8em}  =   \tg(\hxbf^{(j)}) \! + \! L\|\Pi_{\Xc}\big(\hxbf^{(j)}\big) \!-\! \Pi_{\Xc}\big(\xbf^{(j)} \! - \! \alpha \nabla_{\xbf}f(\xbf^{(j)}, \! \ybf^{(j)})\big)\|^{2} \nn\\[-.3em]
& \hspace*{-.8em}  \stackrel{(a)}{\le}  \tg(\hxbf^{(j)}) \! + \! L\|\hxbf^{(j)}\!-\!\xbf^{(j)}\!+ \!\alpha \nabla_\xbf f(\xbf^{(j)}\!,\! \ybf^{(j)})\|^2  \nn \\
& \hspace*{-.8em}  \le \tg_{\frac{1}{\!2L}}\!(\xbf^{(j)}) \!+\! 2\alpha L\nabla_{\xbf}f(\xbf^{(j)}\!, \!\ybf^{(j)})^T\!(\hxbf^{(j)} \!\!- \!\xbf^{(j)}) \!+ \! \alpha^2\!LM^{2}  \label{tg}
\end{align}
where $(a)$ is due to  $\|\Pi_{\Xc}(\xbf_{1}) - \Pi_{\Xc}(\xbf_{2})\|\leq \|\xbf_{1} - \xbf_{2}\|$, $\forall~ \xbf_{1},\xbf_{2}$.
From \eqref{eq_g_ineq},  the second term in \eqref{tg} is given by\vspace{-.4em}
\begin{align}\label{eq_gradt_Moreau}
&\! \nabla_{\xbf}f(\xbf^{(j)}\!, \ybf^{(j)})^T(\hxbf^{(j)} \!- \!\xbf^{(j)}) \nn\\[-.5em]
&\!\!\! \le  \tg(\hxbf^{(j)}) \!- \!\tg(\xbf^{(j)}) \!+ \! \frac{L}{2}\|\hxbf^{(j)} \! - \!\xbf^{(j)}\|^{2} \nn \\
&\!\!\! = \big(\tg(\hxbf^{(j)})\! + \!L\|\hxbf^{(j)}\! - \!\xbf^{(j)}\|^{2}\big) - \!\big(\tg(\xbf^{(j)})\! + \!L\|\xbf^{(j)}\! - \!\xbf^{(j)}\|^{2}\big)\nn\\[-.3em]
& \quad  -{\frac{L}{2}}\|\hxbf^{(j)}\! - \!\xbf^{(j)}\|^{2} \nn\\[-.5em]
&\!\!\! \stackrel{(a)}{\le}   -L\|\hxbf^{(j)}\! - \!\xbf^{(j)}\|^{2} \stackrel{(b)}{=}  -\frac{1}{4L}\|\nabla \tg_{\frac{1}{2L}}(\xbf^{(j)})\|^{2}
\end{align}
where $(a)$ is because $\varphi(\xbf) \triangleq \tg(\xbf) + L\|\xbf - \xbf^{(j)}\|$ is an $L$-strongly convex function,
which leads to\vspace{-.5em}
\begin{align}
&\varphi(\xbf^{(j)})-\varphi(\hxbf^{(j)})=\varphi(\xbf^{(j)})-\min_{\xbf}\varphi(\xbf)  \nn \\[-.5em]
&\!\! \ge \! \nabla\varphi(\hxbf^{(j)})^{T}\!(\xbf^{(j)} \!- \!\hxbf^{(j)}) \! + \! \frac{L}{2}\|\hxbf^{(j)}\! - \!\xbf^{(j)}\!\|^{2} \! = \! \frac{L}{2}\|\hxbf^{(j)}\! - \!\xbf^{(j)}\!\|^{2}  \nn
\end{align}
where the last equality is because $\nabla\varphi(\hxbf^{(j)})=0$ as $\hxbf^{(j)}=\arg\min_{\xbf} \varphi(\xbf)$. Also, $(b)$ in  \eqref{eq_gradt_Moreau} is by \eqref{Moreau_x}.
Applying \eqref{eq_gradt_Moreau} to \eqref{tg} yields\vspace{-.5em}
\begin{align}\label{ineq_diff_gradt}
\!\!\!\tg_{\frac{1}{2L}}(\xbf^{(j+1)}) \le \tg_{\frac{1}{2L}}(\xbf^{(j)}) \!- \!\frac{\alpha}{2}\|\nabla \tg_{\frac{1}{2L}}(\xbf^{(j)})\|^{2} \!+ \!  \alpha^{2}LM^{2}.
\end{align}
Summing both sides of \eqref{ineq_diff_gradt} over $j$, rearranging the terms,
and noting from \eqref{Moreau} that $ \tg_{\frac{1}{2L}}(\xbf^{(J+1)})\ge \min_{\xbf}{\tg(\xbf)}$, we have\vspace*{-.5em}
\begin{align}
\frac{1}{J+1}\sum^{J}_{j=0}\|\nabla \tg_{\frac{1}{2L}}(\xbf^{(j)})\|^{2} & \leq \frac{2\big(\tg_{\frac{1}{2L}}(\xbf^{(0)}) -  \! \min_{\xbf}{\tg(\xbf)}\big)}{\alpha (J+1)} \nn\\
& \qquad + 2\alpha LM^{2}.
\label{ineq_avrg_gradt_dist}
\end{align}
Note from \eqref{Moreau} that $\tg_{\frac{1}{2L}}(\xbf^{(0)})\leq \min_{\xbf}{\tg(\xbf)} + {L\|\xbf^o - \xbf^{(0)}\|^{2}}$, where $\xbf^o \triangleq \arg\min_{\xbf}\tg(\xbf)$. Thus,
$ \tg_{\frac{1}{2L}}(\xbf^{(0)}) - \min_{\xbf}{\tg(\xbf)} \leq{L\|\xbf^o - \xbf^{(0)}\|^{2}} \! \leq \! LD^{2}$.
Also, since  $g(\xbf) = \max_{y\in\Yc}f(\xbf,\ybf)$, $g(\xbf)$ is also $C$-Lipschitz over $\Xc$, \ie $|g(\xbf_1)-g(\xbf_2)| \le C\|\xbf_1-\xbf_2\|$, $\forall \xbf_1,\xbf_2 \in \Xc$. It follows that
$\tg_{\frac{1}{2L}}(\xbf^{(0)}) -  \! \min_{\xbf}{\tg(\xbf)}\le \tg(\xbf^{(0)}) -  \! \min_{\xbf}{\tg(\xbf)} \le CD$.
Combining the above, let $\Delta \triangleq \min{\{LD^{2}, CD\}}$. Then,  \eqref{ineq_avrg_gradt_dist} becomes\vspace{-.5em}
\begin{align}
\frac{1}{J+1}\sum^{J}_{j=0}\|\nabla \tg_{\frac{1}{2L}}(\xbf^{(j)})\|^{2} \leq \frac{2\Delta}{\alpha (J+1)} + 2\alpha LM^{2}.
\label{ineq_avrg_gradient}
\end{align}
Minimizing  RHS of \eqref{ineq_avrg_gradient} over $\alpha$
yields the optimal step size $\alpha = \sqrt{\frac{\Delta}{LM^{2}(J+1)}}$.
Substituting this optimal $\alpha$ into  \eqref{ineq_avrg_gradient} and
noting that  LHS of \eqref{ineq_avrg_gradient} is
$E\|\nabla{\tg_{\frac{1}{2L}}(\bar{\xbf})}\|^{2}$, we have \eqref{ineq_avrg_epsilon}.
\endIEEEproof

\balance

\end{document}